\documentclass{article}

\usepackage{PRIMEarxiv}
\usepackage[colorlinks=true, allcolors=blue]{hyperref}
\usepackage{natbib}
\usepackage{graphicx}
\usepackage{txfonts}
\usepackage[table]{xcolor}
\usepackage{colortbl}
\usepackage{multirow}

\definecolor{lightblue}{rgb}{0.8, 0.85, 1}
\definecolor{lightgreen}{rgb}{0.8, 1, 0.8}
\definecolor{lightyellow}{rgb}{1, 1, 0.8}
\definecolor{lightpink}{rgb}{1, 0.8, 0.8}
\definecolor{lightgray}{rgb}{0.9, 0.9, 0.9}

\title{Solar cycle evolution of ICME sheath regions at 1 AU
\thanks{\textit{\underline{Citation}}: 
\textbf{Larrodera, C., Temmer, M., Owens, M. (2025). “Solar cycle evolution of ICME sheath regions at 1 AU”.
In: Astronomy \& Astrophysics}} 
}

\author{
  C. Larrodera \\
  University of Alcalá \\
  Alcalá de Henares\\
  \texttt{carlos.larrodera@edu.uah.es} \\
  \And
 M. Temmer  \\
 Institute of Physics, University of Graz \\
  Graz\\
  \And
  M. Owens \\
  Space and Atmospheric Electricity Group, Department of Meteorology \\
  Reading\\
}

\begin{document}
\maketitle

\begin{abstract}
We investigate the evolution of interplanetary coronal mass ejection (ICME) sheath regions at 1 AU across solar cycles 23, 24, and the rising phase of 25, focusing on their variability and turbulence in relation to upstream solar wind conditions and the global heliospheric state. Using a dataset of over 900 ICME sheath events, we apply statistical metrics such as the inter-quartile range (IQR) and the Turbulence Index (TI) to quantify variability and turbulence. The analysis compares full and rising phases of solar cycles and examines both local ICME parameters (e.g., sheath total pressure, non-radial flows) and global interplanetary indicators such as open solar flux (OSF). From solar cycle 23 to solar cycle 24, sheath total pressure and magnetic field strength decreased by over 40\% and 25\%, respectively, accompanied by reduced turbulence and variability. In contrast, the rising phase of solar cycle 25 shows increased magnetic complexity, particularly in non-radial field components, despite stable bulk parameters. Non-radial flow patterns also shift from tangentially dominated in solar cycle 23 and solar cycle 24 to normal-dominated in solar cycle 25, suggesting changes in ICME orientation and sheath formation mechanisms. No significant correlation is found between OSF and sheath properties, indicating that local solar wind and ICME-specific factors are the primary drivers of sheath evolution. The study reinforces the importance of upstream solar wind dynamics in relation to variations in plasma and magnetic field measured components of ICME sheaths. The derived trends in turbulence, magnetic orientation, and flow geometry suggest that sheath regions are sensitive indicators of solar cycle phase and should be considered as distinct, structured components in ICME modeling.
\end{abstract}

% keywords can be removed

\section{Introduction}

Coronal mass ejections are solar eruptions of plasma and magnetic fields. The interplanetary counterpart of coronal mass ejections (ICMEs) are detected from in-situ measurements but also observed in heliospheric image data. Typically, we differentiate between three different parts within an ICME: (1) the shock, characterized as a sudden increase in all plasma and magnetic field parameters, (2) the sheath as a compressed plasma region with strongly fluctuating magnetic field and (3) the magnetic obstacle or flux rope identified by low density and temperature, along with smooth magnetic field variation and rotation \citep{Burlaga+1981,Klein+1982,Richardson+1995}.

Compared to the flux rope, the sheath is typically denser and hotter, due to shock compression, and the magnetic field is typically more turbulent and variable \citep[e.g.,][]{Kilpua+2017,Luhmann+2020}. Although conditions at L1 are often assumed to represent those at Earth, recent observations indicate that CME sheath turbulence can evolve significantly even over this short distance \citep{Argall+2025}. As part of the sheath region, situated directly upstream of the magnetic driver, a distinct non-expanding region of enhanced density is identified, presumably corresponding to piled-up plasma \citep{TB22}. \cite{Temmer+2021} found a strong correlation between the sheath density and the upstream solar wind conditions, particularly the solar wind speed and density measured 24 hours before the arrival of the disturbance. Additionally, \cite{Salman2024} found that magnetic field fluctuations within ICME sheaths are strongly linked to the gradient in magnetic field strength between the upstream solar wind and the magnetic ejecta. Hence, the upstream solar wind plays a major role in the development of the sheath and its characteristics. 

The relative speed between the ICME and the local upstream solar wind is related to shock generation, whereas the expansion behavior of the magnetic ejecta influences the characteristics of the sheath structure \citep{Salman+2021}. The magnetic field and speed of the ICME are also correlated as a result of the upstream solar wind compression in the sheath region \citep{Owens+2005}. Distance-dependent relationships have been identified between the behavior of the magnetic ejecta and the properties of the sheath \citep{Larrodera+2024}. During the evolution of the ICME, plasma accumulates along the radial direction of its expansion. However, the expansion and interaction between the solar wind and the interplanetary magnetic field (IMF) also generates non-radial flows (NRFs) within the sheath. Related with the NRFs, \citet{Owens+2004} studied the plasma flows in the non-radial direction induced in the solar wind by the motion of ICMEs, that must arise as fast-moving ICMEs push solar wind plasma aside. \citet{Martinic+2022, Martinic+2023} analyze how the orientation of the ICME affects the NRF and how this will relate to the draping of the IMF around the ICME. They obtained that high-inclined ICMEs have $NRF>1$, therefore the NRF is dominated by the tangential component producing a longitudinal draping. The relevance of the NRF is even more related with the sheath. \citet{AlHaddad+2022} analyze the NRF through different ICMEs obtained from STEREO to obtain that the sheaths are associated with the largest deflection flows. \cite{Jones2002} report about possible effects from the shock orientation on the formation of planar magnetic structures in the sheath, which may facilitate the compression and amplification of magnetic fields in the sheath region. Such planar magnetic structures are found to affect the geomagnetic effectiveness of ICME sheaths regions \citep{palmerio16,lugaz16}. They might also play an important role in particle acceleration processes. 

The ICME rate is not constant and is related to the solar activity and therefore depends on the solar cycle phase. It should be also noticed that the activity of each solar cycle is different. Solar cycle 24 (SC24) revealed much lower activity as compared with the previous solar cycles 21--23. \citet{Gopalswamy+2015} showed that the magnetic obstacles during solar cycle 23 (SC23) had greater radial extents than in SC24. The increased ICME size may be explained by a drop of $\sim$50\% in the sheath total pressure in the heliosphere reported for SC24 \citep[see e.g.,][]{McComas2013}. Differences in the expansion behavior directly influences the ability of ICMEs to drive shocks \citep{Lugaz2017}. In a recent study, \citet{Liu+2025} compared active region areas and magnetic flux, finding a significant decrease during SC24. This highlights the strong connection between surface characteristics and ICME properties. 

In that respect, the open solar magnetic flux (OSF) is a measure of the Sun’s magnetic field that extends outward into the heliosphere. The OSF connects solar surface structures to the IMF, and it therefore plays a crucial role in governing the structure of the heliosphere. Variations in the OSF are closely linked to the evolution of the photospheric magnetic field, which is traced, e.g., by sunspots, as well as by CMEs that can carry OSF away into the heliosphere \citep{FiskSchwadron2001}. As the CME eruption is accompanied by a major restructuring of the coronal magnetic field, previously closed fields may be stretched out into interplanetary space or reconnect, hence, contribute to the OSF. \cite{Owens+2006} and \cite{Owens+2011} proposed an OSF estimation method that includes both a constant component and a non-constant component derived from the CME rate. This results in a sinusoidal evolution of the OSF in phase with the solar cycle. Indeed, \cite{Owens+2011} highlight that CMEs are a major source of open flux addition and can dominate the open flux budget during solar maximum. While the interactions between the magnetic ejecta (flux ropes) of CMEs and the OSF are well covered by studies, there are no studies that investigate the CME sheath region, which is most directly connected to the background solar wind, and the OSF. 

This study focuses on the characteristics of ICME sheath regions in relation to upstream solar wind conditions and the overall state of the heliosphere. We place the sheath in the context of the ambient solar wind, examining how it reflects local and global heliospheric conditions. In particular, we analyze the OSF as a proxy for large-scale heliospheric structure and sheath total pressure, as well as sheath-specific features like non-radial flows. While OSF was initially considered a potential driver of sheath variability, our results show no significant correlation between OSF (excluding ICME contributions) and sheath properties. This indicates that sheath variability is primarily driven by local solar wind and ICME-specific factors. Consequently, our goal goes toward understanding how sheath properties evolve across solar cycle phases. For this purpose, we use the extensive dataset provided by \citet{Larrodera+2024} to conduct a detailed statistical and comparative analysis of these parameters over multiple solar cycles. 

The structure of this article is organized as follows: Section~\ref{sec:data} introduces the ICME sample used in this study, along with the statistical metrics applied throughout the analysis to ensure consistency and comparability. Section~\ref{sec:solar_wind} describes the solar wind conditions preceding the ICME events, focusing on key variables such as the OSF and includes a statistical breakdown to highlight temporal variations. Section~\ref{sec:sheath} characterizes the sheath regions associated with ICMEs by analyzing a set of physical parameters through statistical methods, aiming to identify patterns and anomalies across different solar cycles. Section~\ref{sec:turbulence} details the methodology employed to quantify turbulence within the sheath regions, emphasizing how these turbulent properties evolve over time and differ between solar cycles. Section~\ref{sec:discussion} synthesizes the findings from the sheath analysis, interpreting their implications in the context of solar-terrestrial interactions and space weather forecasting. Finally, Section~\ref{sec:conclusions} summarizes the main conclusions drawn from the study, highlighting the contributions to our understanding of ICME dynamics and suggesting directions for future research.

\section{ICME Data and Statistical Methods} \label{sec:data}

We have selected ICMEs from \cite{Larrodera+2024} covering the distance range between 0.95 AU and 1.05 AU for the time range 1994--2022. More precisely we are using the sheath of these ICMEs understood as the region between the shock and the magnetic obstacle. Our event selection is therefore focused on clear sheath onsets, which may exclude weaker ICMEs. In that regard we also would like to add, that the measured local in-situ signatures depend on whether the spacecraft crosses near the CME nose or flank, and on the background solar wind conditions. While the criterion of a clear sheath ensures a consistent identification of sheath boundaries, it may introduce a bias toward fast and apex-hit ICMEs, which is mitigated by having a large statistical sample.

The data availability spans from the two full solar cycles 23 (SC23; August 1996--December 2008; maximum November 2001) and 24 (SC24; December 2008--December 2019; maximum April 2014), as well as the rising phase of solar cycle 25 (SC25; start December 2019). We note that we discarded solar cycle 22 as only 12 ICME sheath events were detected during the available time range 1994--1996. Table~\ref{tab:sheath_solar_cycle_fps} shows the number of detected ICME sheaths from in-situ measurements for each solar cycle at 1 AU. For a better comparison to solar cycle 25 (SC25), where only the rising phase is available, we perform a two-step analysis. In the first step we analyze data using the full cycle sample (FCS) from SC23 and SC24. In the second step we analyze data covering the rising phase sample (RPS) of SC23, SC24 and SC25 only. The rising phases of SC23 and SC24 are fully covered, while those from SC25 are only partly covered. Table~\ref{tab:sheath_solar_cycle_rps} details the sheaths at 1AU for each rising phase. 

\begin{table}[!htbp]
\caption{Sheaths at 1 AU of the full cycle sample (FCS)} 
\centering                        
\begin{tabular}{|c| c | c|}  
\hline             
&\textbf{Period}& \textbf{\# Sheaths}  \\ \hline
\textbf{SC23} & 01/08/1996 $<$ Date $\leq$ 01/12/2008 & 416 \\ \hline
\textbf{SC24} & 01/12/2008$<$ Date $\leq$ 01/12/2019 & 469  \\ \hline
\textbf{SC25} & Date $>$ 01/12/2019 & 83 \\ \hline
\end{tabular}
 
\label{tab:sheath_solar_cycle_fps}    
\end{table}

\begin{table}[!htbp]
\caption{Sheaths at 1 AU of the rising phase sample (RPS)} 
\centering                        
\begin{tabular}{|c| c | c|}   
\hline             
&\textbf{Period}& \textbf{\# Sheaths}  \\ \hline
\textbf{SC23} & 01/08/1996 $<$ Date $\leq$ 01/11/2001 & 231 \\ \hline
\textbf{SC24} & 01/12/2008$<$ Date $\leq$ 01/04/2014 & 306  \\ \hline
\textbf{SC25} & Date $>$ 01/12/2019 & 83 \\ \hline
\end{tabular}            
\label{tab:sheath_solar_cycle_rps}     
\end{table}

The statistical results obtained from the solar cycle evolution analysis for OSF and ICME sheaths will be shown using boxplots. In the boxplots, the horizontal line represents the median value. The size of the box is defined by the inter-quartile range (IQR) Q$_{3}$-Q$_{1}$. The whiskers extend 1.5 times from the limits of the box, i.e. Q$_{1}$ and Q$_{3}$, beyond which lie the outliers, shown as additional dots. The IQR, as a measure of central data spread, serves as a useful indicator of the sample's statistical robust measure of variability. A low IQR suggests that the data are more clustered around the median, suggesting reduced dispersion. Given its robustness to outliers, the IQR provides a reliable and interpretable metric for assessing variability without being overly influenced by extreme values.
The boxplots and the median values are presented in two different color combinations for FCS and RPS. The FCS will be detailed with a blue box and a red straight line for the median, meanwhile, the RPS will be detailed with a black box and an orange straight line for the median. 

\section{Global heliospheric characteristic conditions} \label{sec:solar_wind}

In this section, we analyze for FCS and RPS the global heliospheric conditions using the OSF. 

\begin{figure}[!htbp]
    \centering
    \includegraphics[width=\columnwidth]{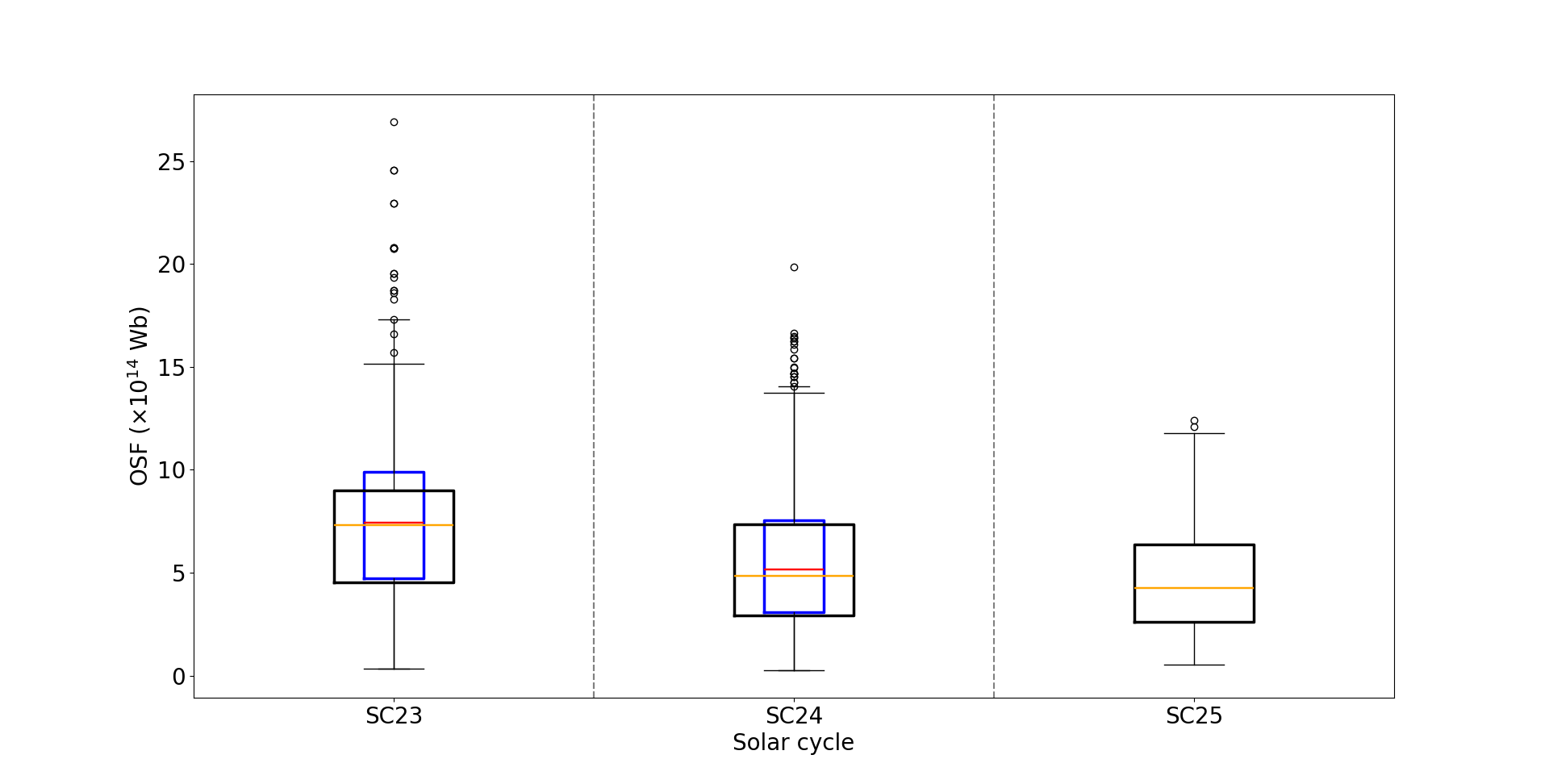}
    \caption{OSF solar cycle boxplot. The FCS box is colored blue with the median line in red and for the RPS the box is colored black with the median line orange.}
    \label{fig:osf_sc_evol}
\end{figure}

The OSF can be understood as the magnetic flux threading the solar photosphere that extends in the heliosphere. We use a similar dataset as described in \citet{Frost+2022} estimating the OSF at 1 AU from in-situ measurements of ACE and WIND between 1994 and 2022. 
However, unlike \citet{Frost+2022}, where ICMEs were not excluded, we remove ICME periods using the time ranges from the Richardson \& Cane ICME list \citep{ICME_catalog}. Therefore, our OSF dataset only covers solar wind contributions. Figure~\ref{fig:osf_sc_evol} shows the distribution of the OSF for the different solar cycles using the FCS and separately for the RPS. The quartile values are given in Table~\ref{tab:osf_sc_evol}. FCS shows between SC23 and SC24 a decrease of the median value of $-$31\%. The IQR, understood as the data spread, decreases by -14\% from SC23--SC24, highlighting less variability during SC24.
RPS shows a very similar decrease of $-$34\% from SC23 to SC24 followed by a smaller decrease $-$12\% from SC24 to SC25. On the other hand, the IQR shows almost no change from SC23 to SC24 with a small decrease $-$16\% from SC24 to SC25.

\begin{table}[!htbp]
\caption{OSF solar cycle boxplots results from Figure \ref{fig:osf_sc_evol}. The values within the brackets are [Q$_{1}$, Q$_{2}$ (median), Q$_{3}$]}
\centering
\begin{tabular}{|c|c|c|}
\hline
\multicolumn{3}{|c|}{\textbf{Open solar Flux (OSF)}} \\ \hline
& \textbf{FCS ($10^{14}$ Wb)} & \textbf{RPS ($10^{14}$ Wb)} \\ \hline
\textbf{SC23} & [4.72, 7.43, 9.91] & [4.54, 7.31, 8.98] \\ \hline
\textbf{SC24} & [3.10, 5.15, 7.54] & [2.91, 4.84, 7.34] \\ \hline
\textbf{SC25} & - & [2.63, 4.27, 6.37] \\ \hline
\end{tabular}
\label{tab:osf_sc_evol}
\end{table}

\section{Sheath characteristic conditions} \label{sec:sheath}

In this section we analyze for FCS and RPS the sheath characteristics by using different variables such as the sheath total pressure, non-radial flow speed and magnetic field. The values of each event are derived over the entire ICME sheath region.

\subsection{Sheath total pressure ($p_{T}$) and Total magnetic field (B)}

The sheath total pressure ($p_{T}$) has been obtained as the sum of the thermal ($p_{Th}=nk_{B}T$) and the magnetic pressure ($p_{M}=B^{2}/2\mu_{0}$), where $T$ and $B$ are the mean temperature and mean magnetic field amplitude of the sheath. In order to see which contribution is more relevant at 1 AU, we computed the pressure ratio $p_{Th}/p_{M}$ (plasma-beta) and obtained that around 85\% of the time, $p_{M}$ dominates over $p_{Th}$. A similar result for fast CMEs is derived in the statistical study from \cite{Masias+2016}.

\begin{figure}[!htbp]
    \centering
    \includegraphics[width=\columnwidth]{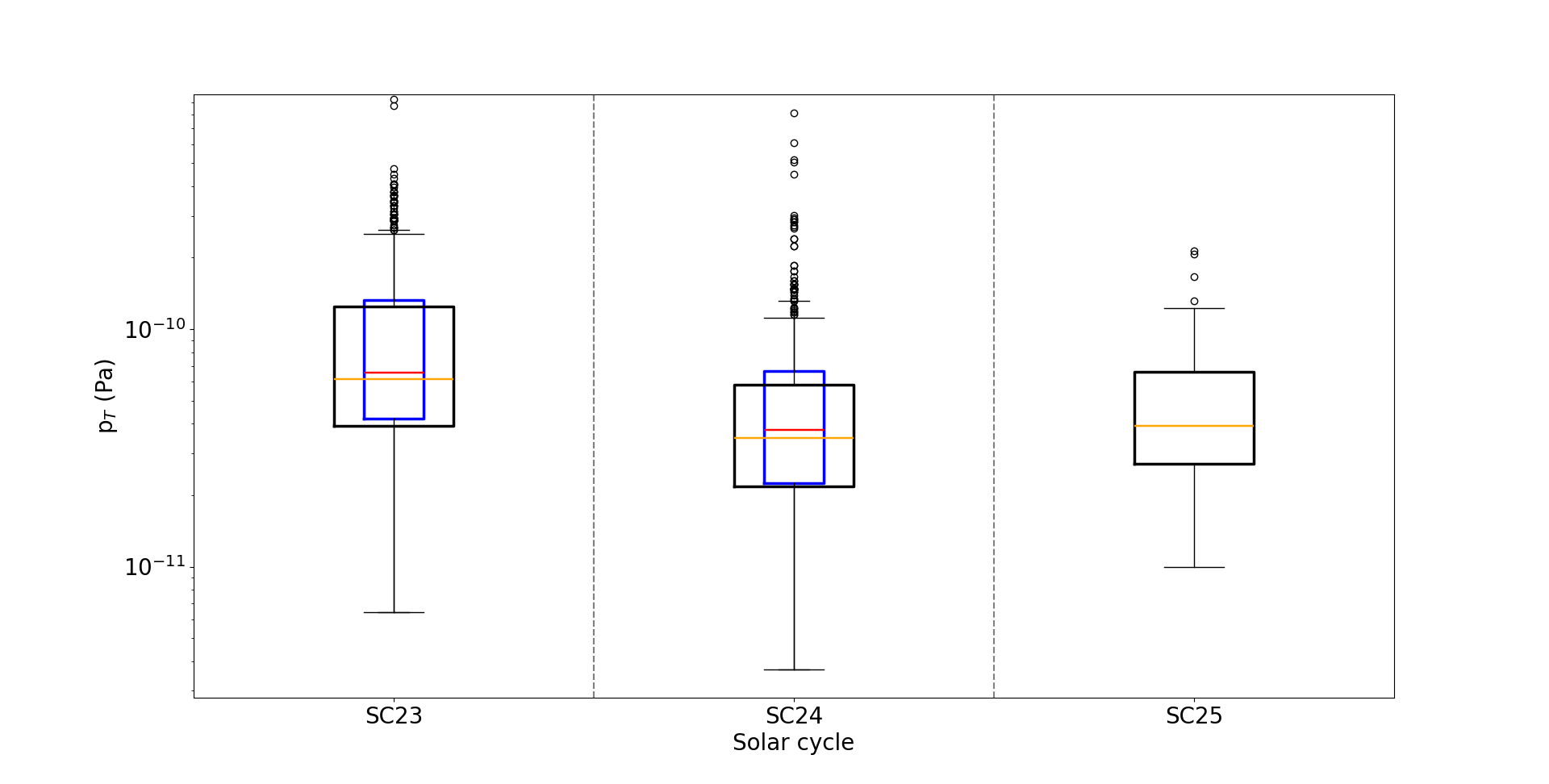}
    \caption{$p_{T}$ solar cycle boxplot. The FCS box is colored blue with the median line in red and for the RPS the box is colored black with the median line orange.}
    \label{fig:pt_sc_evol}
\end{figure}

Figure~\ref{fig:pt_sc_evol} and Table~\ref{tab:pt_sc_evol} show the sheath total pressure ($p_{T}$) boxplot and its quartile values respectively for the FCS and the RPS. For the FCS, SC23 in comparison with SC24 shows a decrease of $-$42\%. The IQR shows a decrease of $-$51\% from SC23 to SC24, i.e. the sheath total pressure drops by half in SC24 as reported by previous studies. The RPS shows for the sheath total pressure, a decrease of $-$43\% from SC23 to SC24, and an increase of +12\% from SC24 to SC25. The IQR shows a large decrease from SC23 to SC24 of $-$57\% and only a +6\% increase from SC24 to SC25. The sheath total pressure in the rising phase of SC25 is comparable to SC24. Hence, compared to SC23 we observe in SC24 and SC25 low pressure and low fluctuations in the sheath.  

As derived before, $p_{T}$ is dominated by $p_{M}$, therefore an increase/decrease in $p_{T}$ is due to an increase/decrease in $B$. Table~\ref{tab:b_sc_evol} shows that indeed, the magnetic field magnitude follows the same trend as $p_{T}$, but the percentages of change are smaller. For the FCS, the median magnetic field decreases from SC23 to SC24 by $-$26\%, with a decrease in data spread, i.e. IQR of $-$36\%. In RPS, the magnetic field median decreases from SC23 to SC24 $-$28\%. The IQR decreases from SC23 to SC24 by $-$42\%. 
These results are aligned with previous authors e.g. \citet{Gopalswamy+2022} which shows SC24 as a weaker solar cycle with $-$32 \% less ICMEs and also less upstream solar wind pressure ($-$17\%) and magnetic field ($-$14\%).
From SC24-SC25, the median magnetic field increase of +10\%, similarly as for the sheath total pressure (+12\%). The only notable difference between the magnetic field magnitude and the sheath total pressure in its trends in RPS is a decrease in the IQR of the magnetic field from SC24 to SC25 ($-$10\%), despite the +6\% increase in sheath total pressure. 

\begin{table}[!htbp]
\caption{$p_{T}$ solar cycle boxplots results from Figure \ref{fig:pt_sc_evol}. The values within the brackets are [Q$_{1}$, Q$_{2}$ (median), Q$_{3}$]} 
\centering                         
\begin{tabular}{|c | c|c|}       
\hline
\multicolumn{3}{|c|}{\textbf{Sheath total pressure ($p_{T}$)}} \\ \hline
& \textbf{FCS ($10^{-11}$ Pa)} & \textbf{RPS ($10^{-11}$ Pa)} \\ \hline 
\textbf{SC23}  & [4.21, 6.54, 13.20] &[3.92, 6.16, 12.50]\\ \hline
\textbf{SC24}  & [2.24, 3.78, 6.64] &[2.17, 3.49, 5.82]\\ \hline
 \textbf{SC25}  & - & [2.72, 3.91, 6.60]\\ \hline
\end{tabular}
\label{tab:pt_sc_evol}     
\end{table}

\begin{table}[!htbp]
\caption{B solar cycle boxplots results. The values within the brackets are [Q$_{1}$, Q$_{2}$ (median), Q$_{3}$]}  
\centering                        
\begin{tabular}{|c | c|c|}       
\hline     
\multicolumn{3}{|c|}{\textbf{Total magnetic field (B)}} \\ \hline
& \textbf{FCS (nT)} & \textbf{RPS (nT)} \\ \hline 
\textbf{SC23}  & [8.01, 10.26, 15.66] &[7.84, 10.17, 14.78]\\ \hline
\textbf{SC24}  & [5.99, 7.64, 10.87] &[5.95, 7.34, 10.0]\\ \hline
 \textbf{SC25}  & - & [6.89, 8.04, 10.52]\\ \hline
\end{tabular}
\label{tab:b_sc_evol}     
\end{table}

\subsection{Non-radial flows and magnetic field orientation}

Within the sheath region, non-radial flows (NRFs) can distort the interplanetary magnetic field (IMF), leading to a draping pattern around the leading edge of the flux rope (driver), rather than the IMF shaping the flow itself \citep{Gosling+1987}. This draping leads to the formation of non-radial flows ($NRFs$) within the sheath \citep{AlHaddad+2022, Martinic+2022}. By analyzing $NRFs$, we can infer the orientation and configuration of the draped IMF. The $NRF$ will also be useful to analyze the different formation mechanism, e.g. propagation or expansion as reported in \citet{Salman+2021}.

\begin{figure}[!htbp]
    \centering
    \includegraphics[width=\columnwidth]{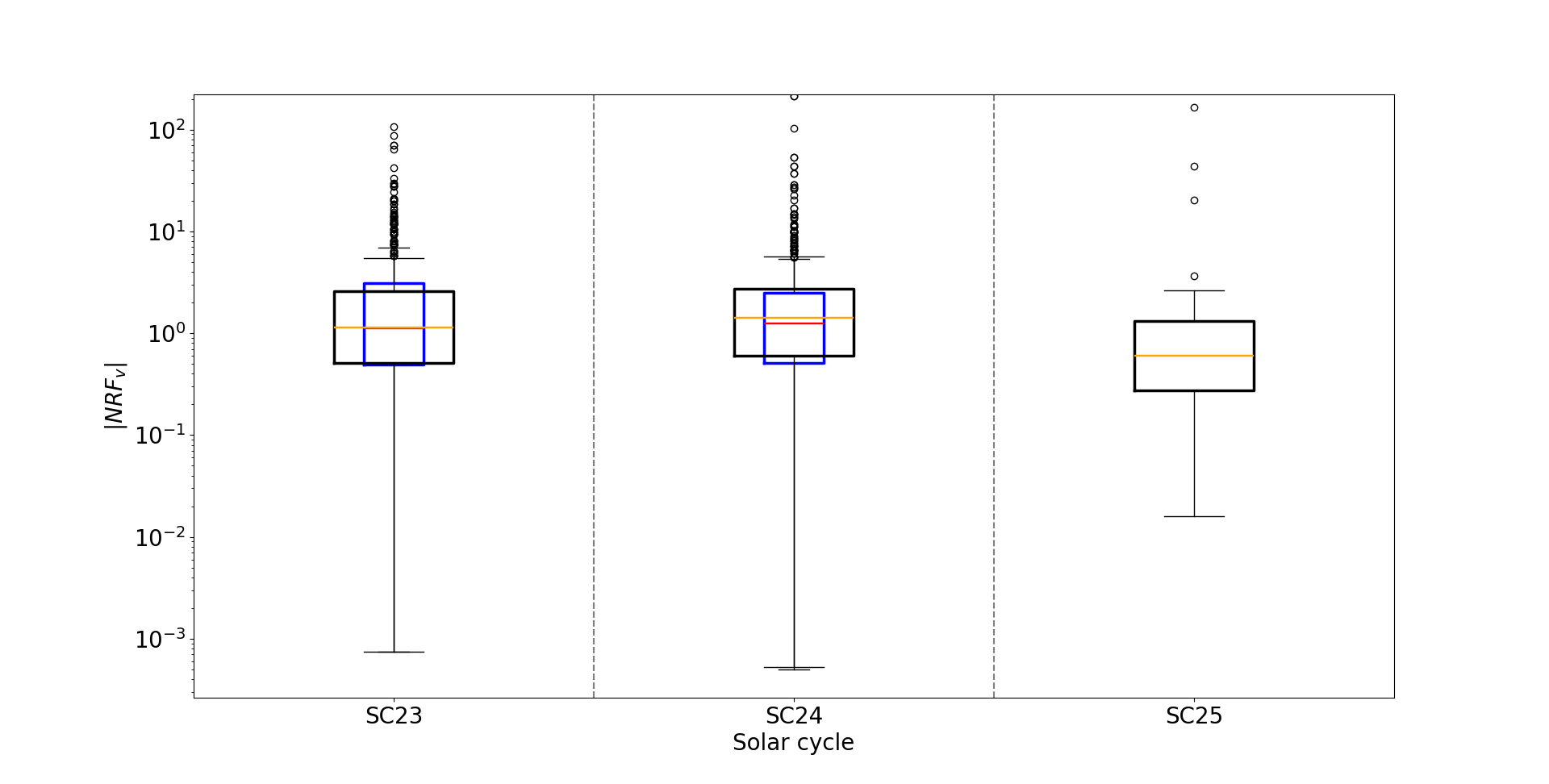}
    \caption{Speed $NRF$ solar cycle boxplot. The FCS box is colored blue with the median line in red and for the RPS the box is colored black with the median line orange.}
    \label{fig:nrf_v_sc_evol}
\end{figure}

For quantification, we decompose the magnetic field and velocity using the RTN reference frame. $NRF$ is defined as the absolute value of the ratio between the tangential and normal components of the velocity, i.e.,  $NRF =|v_T / v_N|$. For the magnetic field, we define the following ratios: $NRB_{TN} = |B_T / B_N|$, $NRB_{TR} = |B_T / B_R|$, and $NRB_{NR} = |B_N / B_R|$. Figure~\ref{fig:nrf_v_sc_evol} and Table~\ref{tab:nrf_v_sc_evol} show the results for $NRF$. For the FCS, $NRF$ increases by +14\% from SC23 to SC24, indicating a stronger tangential component ($v_T$) in SC24. The IQR decreases by approximately $-$36\%, suggesting reduced variability. In the RPS, $NRF$ increases by +24\% from SC23 to SC24, followed by a $-$57\% decrease in SC24--SC25. In SC23, the tangential and normal speed components differ by 15\% ($NRF = 1.15$), while in SC24, the tangential component becomes more dominant ($NRF = 1.43$, a 40\% increase). In SC25, the normal component dominates by about 60\% ($NRF = 0.61$). The IQR remains similar between SC23--SC24 (+2\% difference) but drops by $-$50\% in SC24--SC25, indicating reduced variations.

\begin{figure}[!htbp]
    \centering
    \includegraphics[width=\columnwidth]{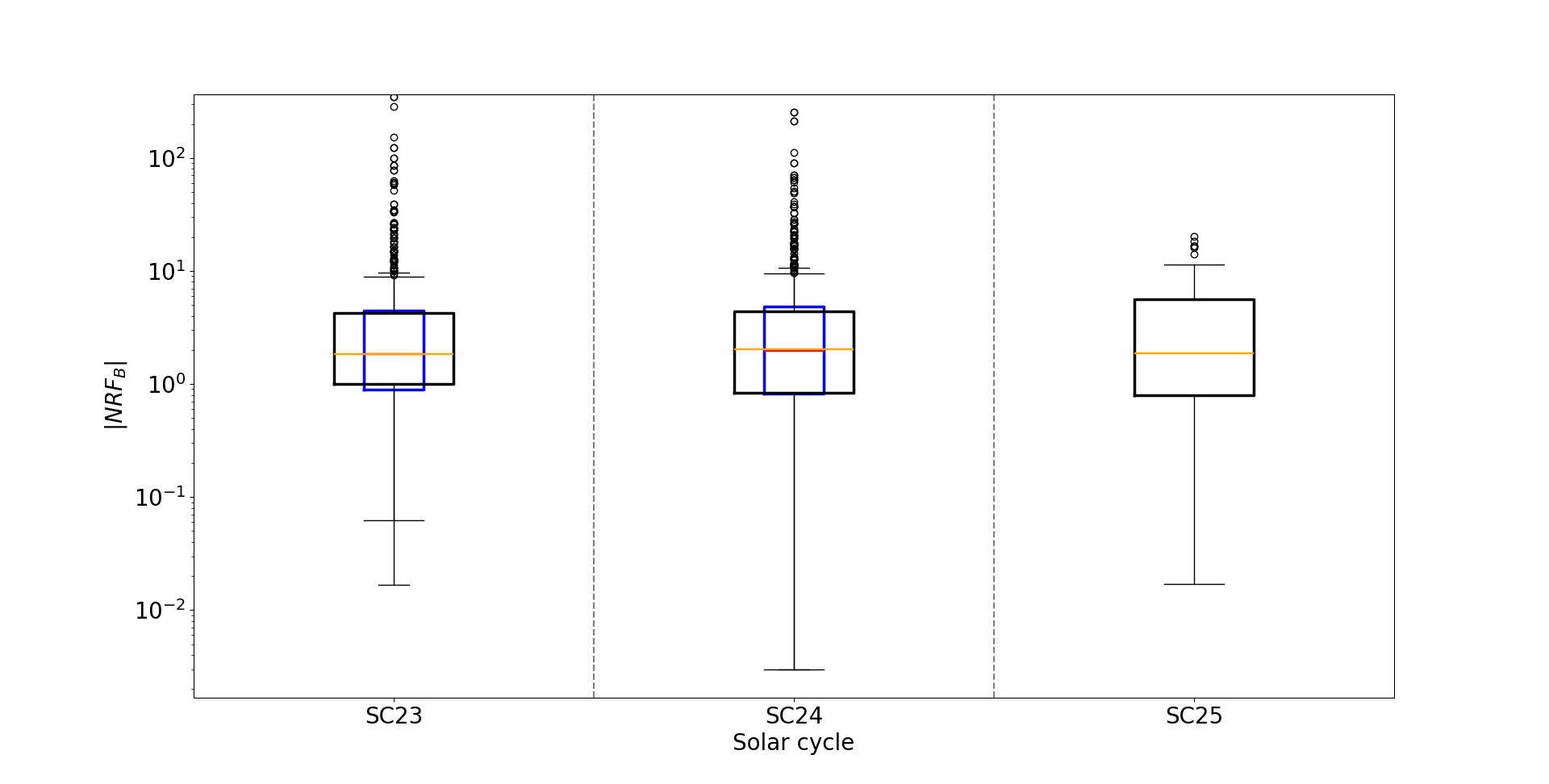}
    \caption{$NRB_{TN}$ solar cycle boxplot. The FCS box is colored blue with the median line in red and for the RPS the box is colored black with the median line orange.}
    \label{fig:nrf_b_sc_evol}
\end{figure}

The results for $NRB_{TN}$ are presented in Figure \ref{fig:nrf_b_sc_evol} and quantified in Table \ref{tab:nrf_b_sc_evol}. It should be noted that the median values of FCS and RPS are $NRB_{TN}>1$, i.e., $B_{T}>B_{N}$, the tangential component is more relevant than the normal component in all the cases. For the FCS, there is a small +6\% increase from SC23 to SC24. The data spread increases +13\% from SC23 to SC24 showing that in this case, SC24 show less dispersion than SC23. For the RPS, we find +11\% increase from SC23 to SC24 followed by a small $-$9\% decrease from SC24 to SC25. In both cases $NRB_{TN}>1$, so the tangential component is more relevant, being even more important in the rising phase of SC24. The data spread shows an increase of +9\% from SC23 to SC24 and +36\% from SC24 to SC25, i.e., the dispersion of the rising phases increases. These results show, as we already see in $NRF$, that in general, the tangential component is more relevant than the normal component. The only exception is for the rising phase of SC25, where it is the opposite.

As previously mentioned, we also computed changes in the magnetic field direction, which we refer to as non-radial orientation of the magnetic field, $NRB_{NR}$ and $NRB_{TR}$. Table~\ref{tab:nrb_combined_structured} presents these quartile values. In the FCS, both $NRB_{TR}$ and $NRB_{NR}$ increase from SC23 to SC24 by +6\% and +9\%, respectively. Since $NRB_{NR}<1$, $B_N$ remains smaller than $B_R$, although the difference narrows in SC24. Conversely, $B_T$ is consistently larger than $B_R$. In the RPS, both components show an upward trend: $NRB_{TR}$ remains stable from SC23--SC24 and SC24--SC25 and increases by +4\% from SC24 to SC25, while $NRB_{NR}$ increases by +8\% and 21\% over the same intervals. Given that sheaths are embedded in the solar wind, their development shape the draping characteristics, hence, are influenced by the IMF. Therefore, we also analyzed the non-radial orientation of the magnetic field in the upstream solar wind. Tables~\ref{tab:nrf_btrupsw_sc_evol} and~\ref{tab:nrf_bnrupsw_sc_evol} summarize these results, with the 'Window' column indicating the duration of the upstream region considered. In the FCS, $NRB_{TR}$ increases from the upstream solar wind to the sheath, regardless of whether a 1-hour or 2-hour window is used. During SC23, $B_T$ is 75\% greater than $B_R$ in the sheath, while in the upstream solar wind this difference ranges from 6\% to 11\%. Combined with the increase in $NRB_{NR}$, indicating a stronger contribution from $B_N$, these results suggest that IMF draping is significant during SC23. The same reasoning applies to SC24, where both $NRB_{NR}$ and $NRB_{TR}$ are higher, indicating more intense draping. Although $NRB_{NR} < 1$, the increasing trend highlights its growing relevance. In the RPS, a similar pattern is observed: for all three rising phases, $NRB_{TR}$ values are lower in the upstream solar wind and increase in the sheath. The same trend is seen in $NRB_{NR}$, which even exceeds 1 during the rising phase of SC25, indicating a stronger non-radial contribution from $B_N$.

\begin{table}[!htbp]
\caption{Speed $NRF$ solar cycle boxplots results from Figure \ref{fig:nrf_v_sc_evol}. The values within the brackets are [Q$_{1}$, Q$_{2}$ (median), Q$_{3}$]. Note that $NRF$ is dimensionless} 
\centering                         
\begin{tabular}{|c| c|c|}     
\hline
\multicolumn{3}{|c|}{\textbf{Non-radial plasma flow (NRF)}} \\ \hline
&  \textbf{FCS} & \textbf{RPS} \\ \hline 
\textbf{SC23}  & [0.50, 1.11, 3.11] & [0.51, 1.15, 2.60]\\ \hline
\textbf{SC24}  & [0.51, 1.26, 2.48]  &[0.60, 1.43, 2.73]\\ \hline
 \textbf{SC25} & -  &[0.27, 0.61, 1.33]\\ \hline
\end{tabular}
\label{tab:nrf_v_sc_evol}     
\end{table}

\begin{table}[!htbp]
\caption{$NRB_{TN}$ solar cycle boxplots results from Figure \ref{fig:nrf_b_sc_evol}.The values within the brackets are [Q$_{1}$, Q$_{2}$ (median), Q$_{3}$]. Note that $NRB_{TN}$ is dimensionless. }
\centering                         
\begin{tabular}{|c| c|c|}       
\hline             
\multicolumn{3}{|c|}{\textbf{Mag. field orientation ($B_{T}/B_{N}$)}} \\ \hline
&  \textbf{FCS} & \textbf{RPS} \\ \hline 
\textbf{SC23}  & [0.89, 1.85, 4.44] & [1.01, 1.84, 4.25]\\ \hline
\textbf{SC24}  & [0.82, 1.96, 4.82]  &[0.84, 2.04, 4.36]\\ \hline
 \textbf{SC25} & -  &[0.80, 1.86, 5.60]\\ \hline
\end{tabular}
\label{tab:nrf_b_sc_evol}      
\end{table}

\begin{table}[!htbp]
\caption{$NRB_{TR}$ and $NRB_{NR}$ solar cycle boxplots results from Figures \ref{fig:nrf_btr_sc_evol} and \ref{fig:nrf_bnr_sc_evol}. The values within the brackets are [Q$_{1}$, Q$_{2}$ (median), Q$_{3}$]. Note that $NRB_{NR}$ and $NRB_{TR}$ are dimensionless. }
\centering
\begin{tabular}{|c|c|c|c|}
\hline
\multicolumn{4}{|c|}{\textbf{Mag. field orientation ($B_{N}/B_{R}$ \& $B_{T}/B_{R}$)}} \\ \hline

\multicolumn{2}{|c|}{} & \textbf{FCS} & \textbf{RPS} \\ \hline

\multirow{2}{*}{\textbf{SC23}} & $NRB_{NR}$ & [0.36, 0.88, 2.14] & [0.40, 0.88, 2.17] \\ \cline{2-4}
                               & $NRB_{TR}$ & [0.90, 1.75, 3.39] & [1.04, 1.93, 3.44] \\ \hline

\multirow{2}{*}{\textbf{SC24}} & $NRB_{NR}$ & [0.42, 0.96, 2.69] & [0.46, 0.95, 2.70] \\ \cline{2-4}
                               & $NRB_{TR}$ & [0.92, 1.85, 3.96] & [0.92, 1.93, 4.13] \\ \hline

\multirow{2}{*}{\textbf{SC25}} & $NRB_{NR}$ & - & [0.47, 1.15, 2.17] \\ \cline{2-4}
                               & $NRB_{TR}$ & - & [0.92, 2.00, 4.36] \\ \hline
\end{tabular}
\label{tab:nrb_combined_structured}
\end{table}

\begin{figure}[!htbp]
    \centering
    \includegraphics[width=\columnwidth]{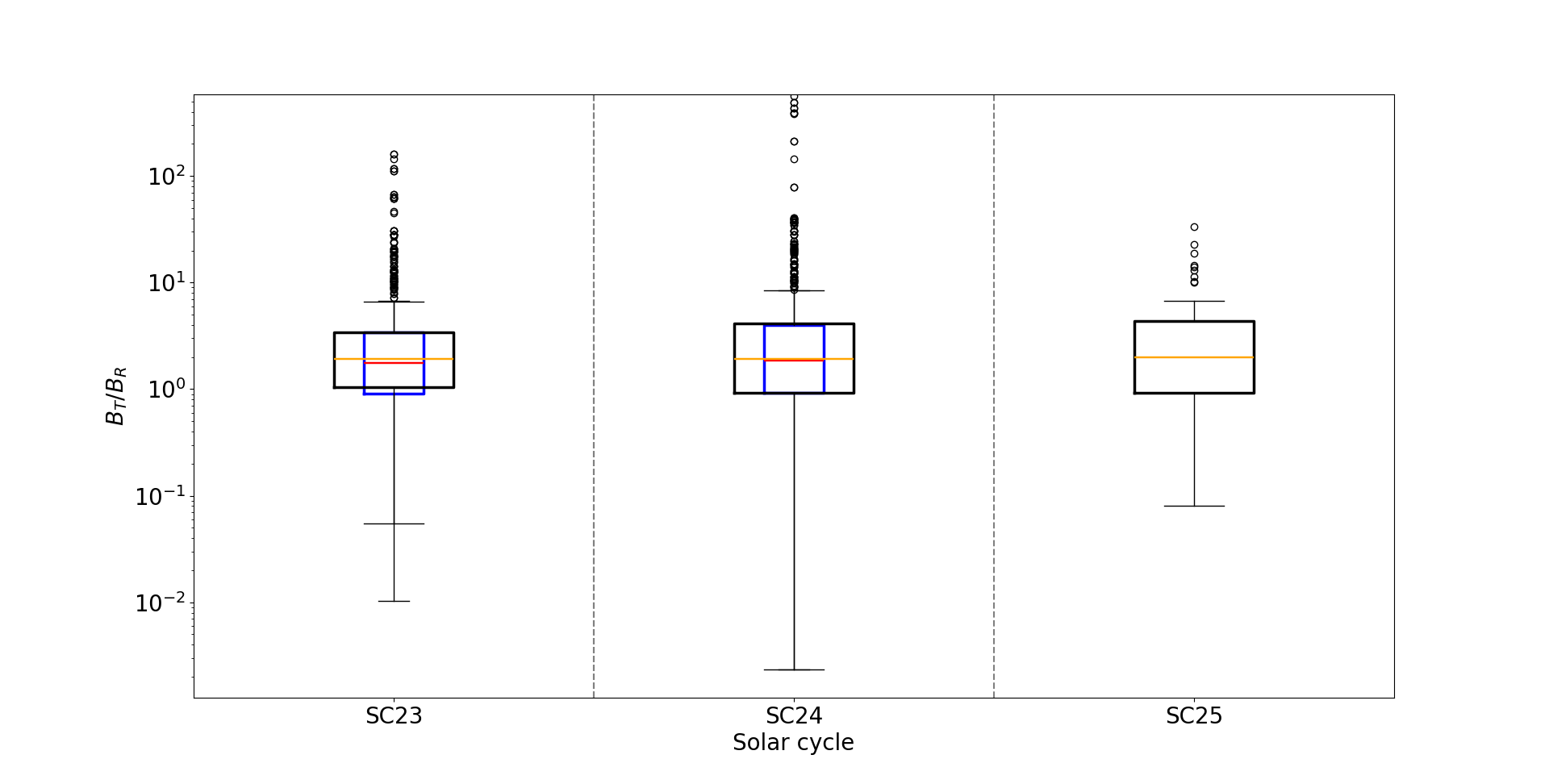}
    \caption{$NRB_{TR}$ solar cycle boxplot. The FCS box is colored blue with the median line in red and for the RPS the box is colored black with the median line orange.}
    \label{fig:nrf_btr_sc_evol}
\end{figure}

\begin{figure}[!htbp]
    \centering
    \includegraphics[width=\columnwidth]{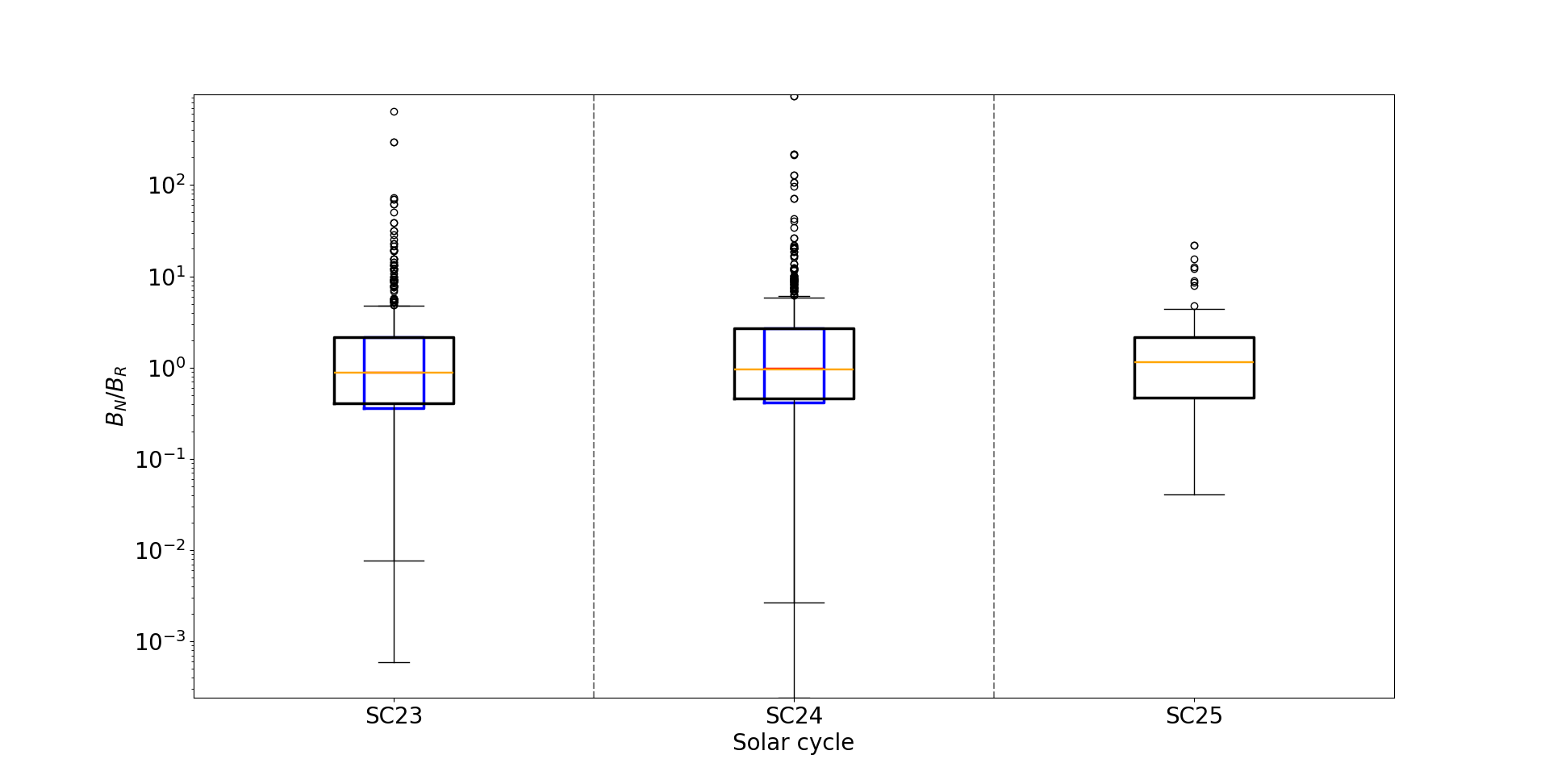}
    \caption{$NRB_{NR}$ solar cycle boxplot. The FCS box is colored blue with the median line in red and for the RPS the box is colored black with the median line orange.}
    \label{fig:nrf_bnr_sc_evol}
\end{figure}

\begin{table}[!htbp]
\caption{$NRB_{TR}$ solar cycle boxplots results of the upstream solar wind. The 'Window size' column indicates the duration of the interval selected just before the sheath region. The values within the brackets are [Q$_{1}$, Q$_{2}$ (median), Q$_{3}$]}
\centering
\begin{tabular}{|c|c|c|c|}
\hline
& \textbf{Window size} &\textbf{FCS} & \textbf{RPS} \\ \hline

\multirow{2}{*}{\textbf{SC23}} & 1h&[0.43, 1.11, 2.43] & [0.56, 1.17, 2.72] \\ \cline{2-4}
                               & 2h&[0.48, 1.06, 2.23] & [0.53, 1.12, 2.29] \\ \hline

\multirow{2}{*}{\textbf{SC24}} & 1h&[0.53, 1.28, 2.86] & [0.53, 1.29, 2.82] \\ \cline{2-4}
                               & 2h&[0.56, 1.22, 3.14] & [0.53, 1.22, 3.14] \\ \hline

\multirow{2}{*}{\textbf{SC25}} & 1h                &-& [0.58, 1.51, 2.87] \\ \cline{2-4}
                               & 2h                & -&[0.53, 1.49, 2.54] \\ \hline
\end{tabular}
\label{tab:nrf_btrupsw_sc_evol}
\end{table}

\begin{table}[!htbp]
\caption{$NRB_{NR}$ solar cycle boxplots results of the upstream solar wind. The 'Window size' column indicates the duration of the interval selected just before the sheath region. The values within the brackets are [Q$_{1}$, Q$_{2}$ (median), Q$_{3}$] }
\centering
\begin{tabular}{|c|c|c|c|}
\hline
& \textbf{Window size} &\textbf{FCS} & \textbf{RPS} \\ \hline

\multirow{2}{*}{\textbf{SC23}} & 1h&[0.30, 0.63, 1.52] & [0.31, 0.68, 1.74] \\ \cline{2-4} 
                               & 2h&[0.26, 0.62, 1.40] & [0.30, 0.64, 1.47] \\ \hline

\multirow{2}{*}{\textbf{SC24}} & 1h&[0.34, 0.75, 1.76] & [0.38, 0.83, 1.89] \\ \cline{2-4}
                               & 2h&[0.26, 0.68, 1.57] & [0.31, 0.76, 1.62] \\ \hline

\multirow{2}{*}{\textbf{SC25}} & 1h                &-& [0.34, 0.81, 2.00] \\ \cline{2-4}
                               & 2h                & -&[0.31, 0.72, 2.42] \\ \hline
\end{tabular}
\label{tab:nrf_bnrupsw_sc_evol}
\end{table}

\section{Quantifying sheath turbulence using TI} \label{sec:turbulence}

The Turbulence Index (TI), defined as the ratio between the IQR and the median (Q$_{2}$), $TI=IQR/Q_{2}$, offers a relative measure of turbulence within a dataset. Unlike absolute dispersion metrics, TI contextualizes the spread of the data in relation to its median value, making it particularly useful for comparing variations across variables with different scales. A lower TI indicates that the turbulence is small relative to the median, suggesting a lower amount of variations in the system. As a robust and dimensionless indicator, TI complements traditional dispersion measures as the IQR, by highlighting proportional changes in data structure that might otherwise go unnoticed.

\subsection{RPS results}

\begin{figure}[!htbp]
    \centering
    \includegraphics[width=\columnwidth]{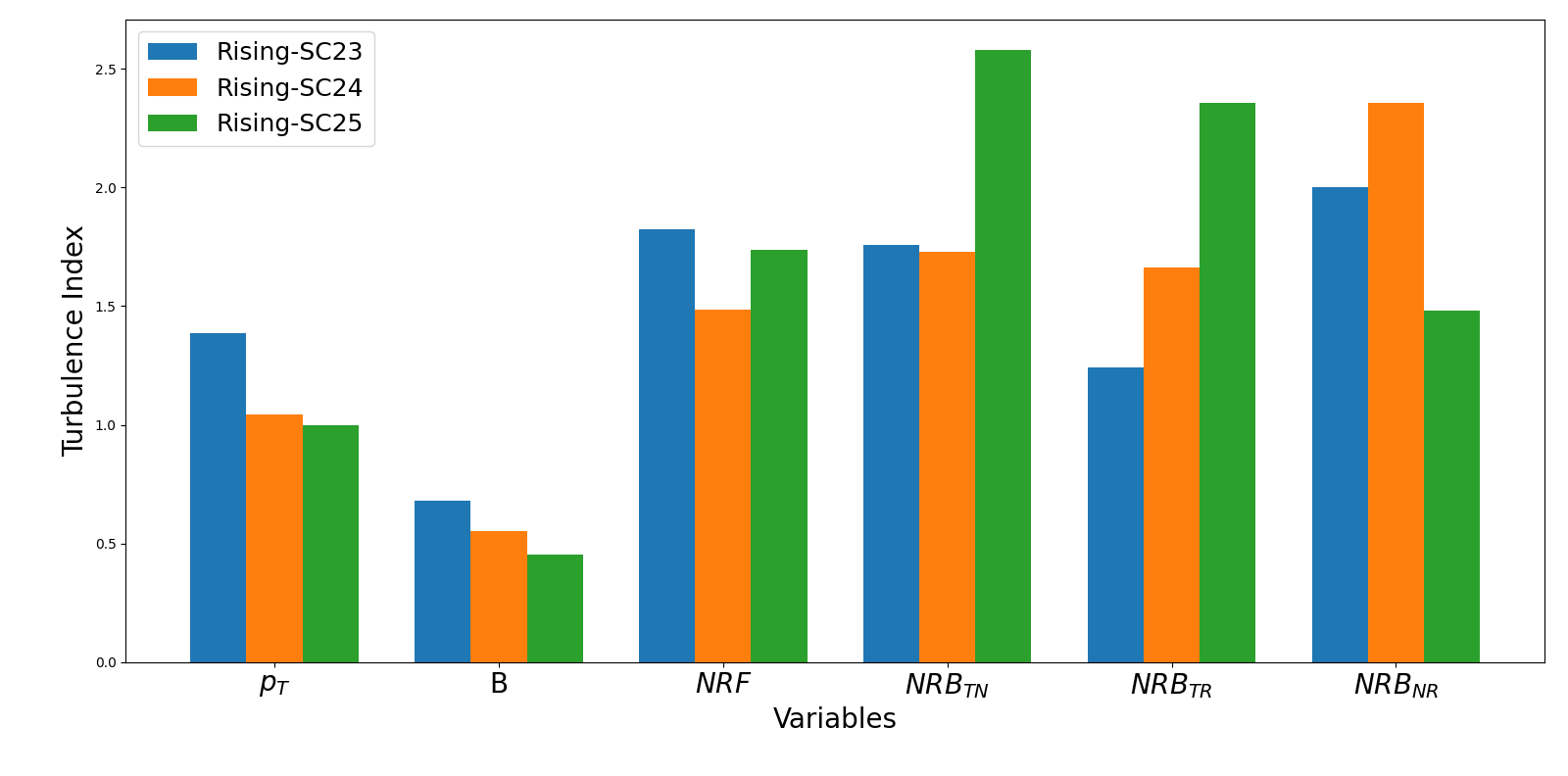}
    \caption{Turbulence Index (TI) for the RPS of the total pressure (p$_T$), non-radial flow speed ($NRF$), magnetic field magnitude (B) and non-radial magnetic field ($NRB_{TN}$, $NRB_{TR}$ and $NRB_{NR}$). The blue, orange and green bars shows the TI values for the rising phase of SC23, SC24 and SC25 respectively.}
    \label{fig:TI_rising_sample}
\end{figure}

Table~\ref{table:TI_rising_sample} details the RPS--TI of the different sheath characteristic variables analyzed: sheath total pressure ($p_{T}$), magnetic field magnitude ($B$), non-radial flow speed ($NRF$) and non-radial magnetic field ($NRB_{TN}$, $NRB_{TR}$ and $NRB_{NR}$). All these have been obtained for the entire sheath region through the median (Q$_2$) and the IQR, as explained before. Complementary, Figure~\ref{fig:TI_rising_sample} shows these same results but with blue, orange and green bars for the RPS of SC23, SC24 and SC25 respectively, showing the TI values obtained for each variable.

The sheath total pressure shows a constant decrease in TI from to SC23 to SC25, being more pronounced from SC23 to SC24. This suggests that over the solar cycles the sheath fluctuations progressively become lower. Since $p_{T}$ is primarily governed by the magnetic field, this trend likely reflects a reduction in magnetic field fluctuations within the sheath. Indeed, from Table~\ref{table:TI_rising_sample}, TI--$B$ has the same trend as $p_{T}$, but in this case the level of fluctuations is smaller. Actually, $B$ shows the lowest TI from all the variables under study. These results reinforce the idea obtained from $p_{T}$, the fluctuation in RPS is reduced from SC23--SC25. The previous section showed that in both cases, the data from $p_{T}$ and $B$ were less spread around the median in the rising phases of SC23 and SC24. This result complements with the fact of less fluctuations in both rising phases. In the rising phase of SC25, $B$ also reduces the spread but in $p_{T}$ shows a small increase.

On the other hand, all the non-radial flows studied have higher TI than $p_{T}$ and $B$. $NRF$ shows a decrease from the rising phase SC23--SC24 followed by almost the same increase in SC25. This reflects the less turbulent rising SC24 phase and it is aligned with what we observed in previous results, i.e., SC24 is the solar cycle with less variability. The non-radial orientation in the magnetic field show different trends. $NRB_{TN}$ and $NRB_{TR}$, both related with the tangential component show that the highest TI is for the rising phase of SC25. Meanwhile, the highest turbulence of $NRB_{NR}$ is for the rising phase of SC24. 

\begin{table}[!htbp]
\caption{Turbulence Index (TI) for the RPS of the total pressure (p$_T$), non-radial flow speed ($NRF$), magnetic field magnitude (B) and non-radial magnetic field ($NRB_{TN}$, $NRB_{TR}$ and $NRB_{NR}$) from Figure \ref{fig:TI_rising_sample}.}
\centering
\begin{tabular}{|c|c|c|c|}
\hline
\multicolumn{4}{|c|}{\textbf{Turbulent index (TI)}} \\
\hline
\textbf{Variable} & \textbf{SC 23} & \textbf{SC 24} & \textbf{SCS 25} \\
\hline
\textbf{p$_T$} & 1.387 & 1.045 & 0.996 \\
\hline
\textbf{$NRF$} & 1.822 & 1.487 & 1.737 \\
\hline
\textbf{B} & 0.682 & 0.551 & 0.453 \\
\hline
\textbf{$NRB_{TN}$} & 1.758 & 1.728 & 2.579 \\ \hline
\textbf{$NRB_{TR}$} & 1.242 & 1.661 & 2.356 \\ \hline
\textbf{$NRB_{NR}$} & 2.001 & 2.356 & 1.483 \\ \hline
\end{tabular}
\label{table:TI_rising_sample}
\end{table}

\subsection{FCS results}

\begin{figure}[!htbp]
    \centering
    \includegraphics[width=\columnwidth]{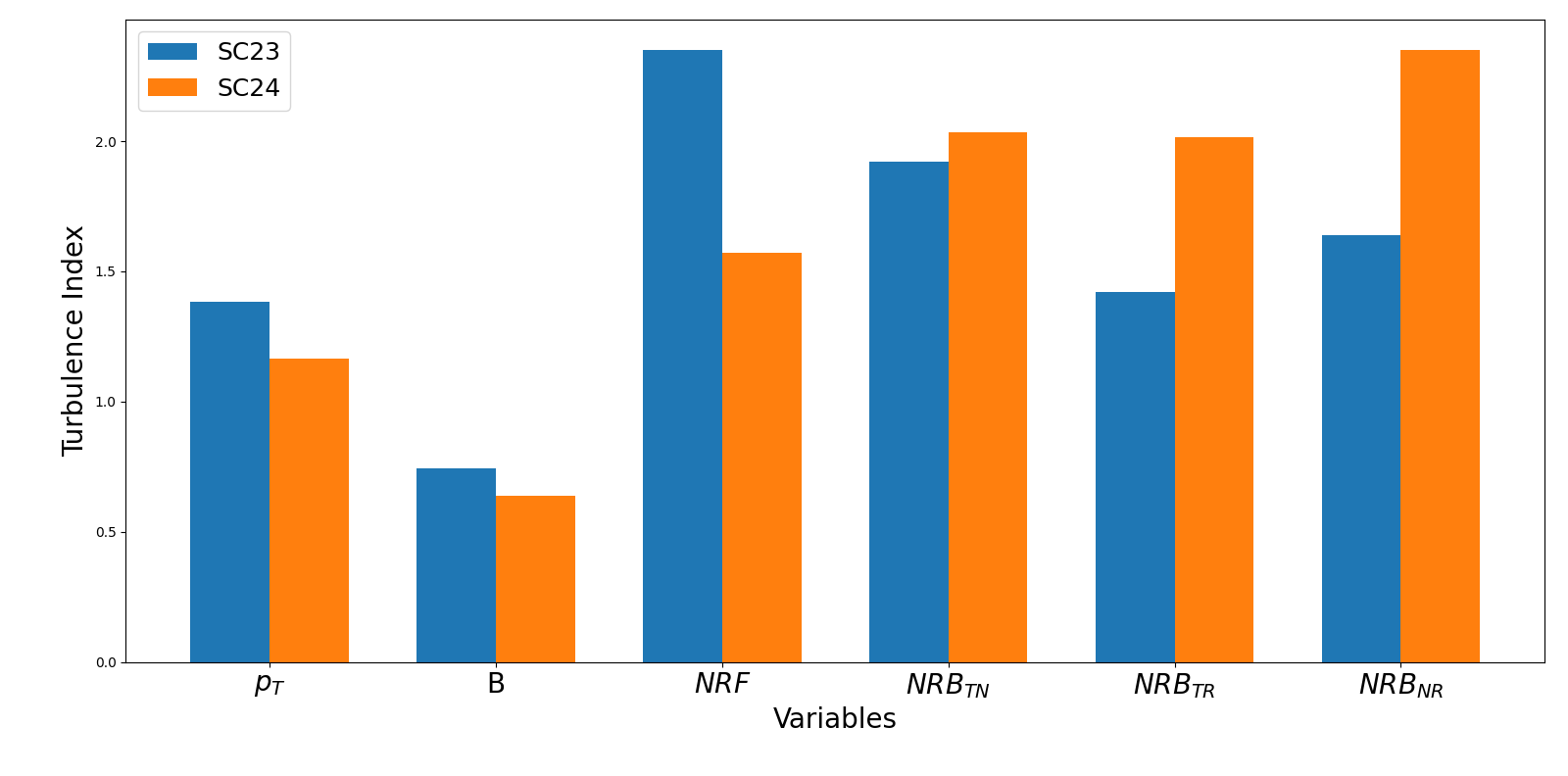}
    \caption{Turbulence Index (TI) for the FCS of the total pressure (p$_T$), non-radial flow speed ($NRF$), magnetic field magnitude (B) and non-radial magnetic field ($NRB_{TN}$, $NRB_{TR}$ and $NRB_{NR}$). The blue and orange bars shows the TI values for the entire solar cycles 23 and 24 respectively.}
    
    \label{fig:TI_sc_sample}
\end{figure}

The FCS-TI values obtained from the different variables analyzed are detailed in Table~\ref{table:TI_sc_sample}, meanwhile  Figure~\ref{fig:TI_sc_sample} shows these TI values for each variable with blue and orange bars for SC23 and SC24 respectively.

SC24 continues to exhibit lower turbulence levels in $p_{T}$ and $B$, consistent with the results obtained from the RPS analysis. This supports the characterization of SC24 as a less intense solar cycle. $p_{T}$ follows the same trend observed in the rising phase, i.e., a decrease in turbulence from SC23 to SC24. This complements the reduced data-spread observed also for SC24 through the IQR in the previous section. As occurred in the FCS, the $B$ turbulence also decreases from SC23 to SC24, although it is smaller than the turbulence of $p_{T}$. In this case, as occurred for the sheath total pressure, the magnetic field magnitude in SC24 has fewer fluctuations than SC23.

The TI-FCS non-radial flow show different trends. On one hand, for $NRF$, SC23 is more turbulent than SC24. On the other hand, all the magnetic field non-radial flows show the opposite. SC24 is more turbulent than SC23. This indicates that even though SC24 was less active overall, the magnetic field turbulences within sheaths became more pronounced. These results collectively suggest that while bulk plasma properties became more homogeneous, the magnetic field configuration within sheaths evolved toward greater complexity. This divergence may reflect changes in magnetic topology or upstream solar wind conditions that disproportionately affect magnetic field behavior.

\begin{table}[!htbp]
\caption{Turbulence Index (TI) for the FCS of the total pressure (p$_T$), non-radial flow speed ($NRF$), magnetic field magnitude (B) and non-radial magnetic field ($NRB_{TN}$, $NRB_{TR}$ and $NRB_{NR}$) from Figure \ref{fig:TI_sc_sample}}
\centering
\begin{tabular}{|c|c|c|}
\hline
\multicolumn{3}{|c|}{\textbf{Turbulent index (TI)}} \\
\hline
\textbf{Variable} & \textbf{SC 23} & \textbf{SC 24} \\
\hline
\textbf{p$_T$} & 1.382 & 1.164 \\
\hline
\textbf{$NRF$} & 2.350 & 1.571 \\
\hline
\textbf{B} & 0.745 & 0.639 \\
\hline
\textbf{$NRB_{TN}$} & 1.923 & 2.036 \\ \hline
\textbf{$NRB_{TR}$} & 1.421 & 2.017 \\ \hline
\textbf{$NRB_{NR}$} & 1.641 & 2.350 \\ \hline
\end{tabular}
\label{table:TI_sc_sample}
\end{table}

\section{Discussion} \label{sec:discussion} 
The sheaths driven by ICMEs are the first structures to interact with Earth's magnetosphere and can cause, especially due to strong fluctuations in the magnetic field, geomagnetic storms independent of the magnetic ejecta \citep{Pulkkinen+2007, Myllys+2016}. To better understand the characteristics of ICME sheath regions, we perform a statistical study and cross-compare events during solar cycles 23 to 25 (SC23--25) in two samples. The first sample involves ICMEs from the full solar cycles (FCS), i.e., SC23 and SC24. The second sample, to account for the missing decline phase in SC25, only ICMEs from the rising phases (RPS) of SC23, SC24 and SC25. To describe the sheath characteristics we derived the sheath total pressure ($p_{T}$), non-radial flow speed ($NRF$) and non-radial magnetic field ($NRB_{TN}$, $NRB_{TR}$ and $NRB_{NR}$). For each sample we calculate the inter-quartile range (IQR) to quantify the overall variability of the variables and the turbulence index (TI) which is a measure of the fluctuations scaled by the median value. To describe the solar wind environment in which the ICMEs travel, and to see how the overall heliospheric magnetic conditions might affect the  sheath regions, we use the upstream magnetic field and the measured open solar flux (OSF) in interplanetary space.

\subsection{Relation to heliospheric conditions}
We find that the sheath total pressure, sheath median magnetic field, and the median open solar flux (OSF) - excluding the ICME contribution - decrease in both the RPS and FCS samples over the solar cycles (see Section~\ref{sec:solar_wind}). Although these trends suggest a possible physical connection between the parameters, the Spearman correlation coefficients between the OSF and the extracted sheath variables show no significant correlation. Therefore, while ICMEs are found to make a substantial contribution to the OSF and can nearly double it from solar minimum to maximum \citep{Owens+2011}, our results indicate that the OSF is not directly related to the sheath properties. The Spearman correlation coefficient calculated between the average sheath magnetic field and the 1 hour (2 hours) upstream magnetic field is around 0.50 in both cases RPS and FCS, showing a moderate correlation. This agrees with the findings of \cite{Salman2024}, who reported a strong correlation (0.76) between the upstream magnetic field fluctuations and the ratio of the magnetic field strength in the magnetic ejecta (ME) to that in the upstream solar wind (SW), $B_{ME}/B_{SW}$ \citep[see also][]{Temmer+2021}. By extending the analysis to all other extracted variables and their possible combinations, we found no significant correlations, indicating the absence of meaningful relationships among the variables examined.

\subsection{Solar cycle variations - amplitudes}
The comparative analysis of the sheath-related variables across the solar cycles is summarized in Table~\ref{tab:summary}. These results reveal a clear transition from SC23 to SC24, characterized by a significant decrease in both $p_T$ (around $\sim$40\%) and $B$ (around $\sim$25\%), accompanied by reduced variability and turbulence. This trend is consistent across both FCS and RPS, supporting the interpretation of SC24 as a period of enhanced sheath stability and reduced heliospheric activity. The TI analysis further confirms this, showing lower turbulence levels in SC24 for most parameters. The RPS for SC24 shows a thermal pressure contribution of 36\% against 29\% for the RPS of SC23. Between SC24 and SC25 the thermal pressure might give a higher contribution to the sheath total pressure. These results are supported by previous studies finding a reduced magnetic field magnitude within ICME's magnetic ejecta and sheaths \citep{Gopalswamy+2014,Yermolaev+2021,Gopalswamy+2022} as well as a reduced number of observed ICMEs during SC24 \citep{Yermolaev+2022b}.

The solar cycle related analysis for the global heliospheric condition using the OSF (see Table~\ref{tab:summary2}) shows similar trends. For the FCS, the median OSF decreases by 30\% from SC23 to SC24 with a reduction in the data spread around the median by 14\%. In the RPS, the OSF reveals a strong decrease ($-$34\%) in the median value for the rising phases from SC23--SC24, followed by a weaker one ($-$12\%) from SC24--SC25. The data spread shows no change from SC23--SC24 with a decrease from SC24--SC25 of $-$16\%. This indicates that a reduction in the OSF is accompanied by a decrease in its complexity, except for the RPS of SC23 and SC24, where there was no change in spread.

\subsection{Solar cycle variations - non-radial flows and magnetic field orientation} 
For the FCS from SC23 to SC24 we find an increase in the $NRF$, hence, an enhancement in the contribution of $v_{T}$. This is accompanied by a decrease in the data spread and TI. The non-radial magnetic field ($NRB_{TN}$, $NRB_{TR}$ and $NRB_{NR}$) behavior is different. While the median values, increase too, they are of smaller amount compared to the $NRF$. However, data spread and TI show enhancements. Concluding, compared to SC23, for SC24 the non-radial plasma flow behavior is less turbulent while the non-radial behavior of the magnetic field turns out to be more turbulent. For the RPS there is a change in $NRF$ relevance, from the dominant component $v_{T}$ in SC23 and SC24, to $v_{N}$ in SC25 accompanied by higher turbulence (see Table \ref{tab:nrf_v_sc_evol}). The non-radial magnetic field ratio $B_{T}/B_{N}$ reveals a dominance of $B_{T}$ for all cycles and rising phases (see Table \ref{tab:nrf_b_sc_evol}). $B_{N}/B_{R}$ shows a dominance of $B_{R}$ over all cycles and rising phases except for SC25 during which $B_{N}$ dominates. The tangential component shows over the radial component a dominance over all cycles and rising phases (see Table \ref{tab:nrb_combined_structured}).

ICME inclination is related to the NRF and will also affect how the IMF drapes. Our results (see Table~\ref{tab:nrf_v_sc_evol}) show that for both cases, FCS and RPS, the derived $NRF$ median value is greater than 1 except for rising phase of SC25. \citet{Martinic+2022} gives for high-inclined ICMEs a mean $NRF$ of 1.50$\pm$0.60, and for low-inclined ICMEs a mean $NRF$ of 0.98$\pm$0.37. Considering the mean value and one standard deviation, an overlap between both regimes can be argued in the range of 0.89 and 1.35. Consequently, the $NRF$ values obtained in our research reveal that only the rising phases of SC24 and SC25 are outside this overlap and can be classified as high and low inclination ICMEs, respectively. All other events occur within this interval near the upper boundary and therefore tend to be classified as highly inclined ICMEs.

\subsection{Relation to local upstream conditions}
The relationship between the upstream and sheath magnetic fields is of particular relevance. For both non-radial magnetic field ratios, $NRB_{TR}$ and $NRB_{NR}$, we observe an enhancement from upstream to sheath part. $NRB_{TR}$ is found in all cycles and phases to be significantly larger than one, hinting towards the influence on the tangential magnetic field component in the sheath formation. This might be expected due to the Parker spiral which is reflected in the dominance of $B_{T}$ over $B_{R}$. $NRB_{NR}$ has values much lower than one for all cycles and phases, with a slight increase from upstream to sheath part. Hence, we might see the intensification of the normal component reflecting the compression process in the sheath, adding out-of-ecliptic components to the magnetic field. This is found to be exceptional striking for SC25, where the sheath $NRB_{NR}$ even has a value larger than one. Combined with the finding that the rising phase of SC25 presumably includes more low-inclination ICMEs, the compression may be more efficient in such a geometry, as the plasma can less easily flow around the extended ICME body.

A rise in $NRB_{TR}$ from upstream to sheath may indicate solar wind deflection at the nose of the ICME. Similarly, an increase in $NRB_{NR}$ from the upstream region to the sheath can be interpreted as a proxy for compression and enhanced plasma pile-up. Both effects are consistent with the sheath formation mechanisms proposed by \citet{Salman+2021}, who suggested (i) a propagation mechanism, in which the solar wind is deflected around the nose of the magnetic obstacle, and (ii) an expansion mechanism, in which the obstacle expands without significant propagation, leading to plasma accumulation ahead of it. In that respect, our results suggest a trend towards a dominance in the propagation sheath formation mechanism. An exception occurs for the rising phase of SC25 where the normal component clearly dominates referring to an effect of expansion. This aligns with the findings of \citet{Larrodera+2024}, who, using a different approach, showed that expansion becomes dominant only in the outer heliosphere, whereas up to 1 AU both propagation and expansion contribute.

\subsection{Limitations}

We emphasize that our analysis interprets non-radial flows within the framework of classical ICME structures, typically represented by a croissant-shaped flux rope expanding self-similarly through interplanetary space \citep[e.g.,][]{Zurbuchen+2006, Thernisien+2009, Wang+2018}. However, many ICMEs deviate from this idealized geometry due to interactions with the ambient solar wind, which can modify the observed magnetic field pile-up and non-radial flows depending on the relative orientation and local conditions. Recent modeling studies \citep[e.g.,][]{Rodriguez+2024, Wyper+2024} indicate that spheromak-like or pseudostreamer-associated CMEs may be relatively common, particularly during weaker solar cycles. Such alternative configurations, together with possible magnetic reconnection between the ejecta and the ambient IMF, introduce additional complexity that can influence the observed flow patterns and magnetic field orientations. While our statistical results reveal consistent trends across solar cycles, this structural diversity should be considered when interpreting individual events or when comparing different heliospheric conditions.

Finally, it should be noted that the characterization of sheath onsets and ICME structures is inherently influenced by the spacecraft’s position relative to the ICME and the ambient solar wind. The local in-situ signatures, used to extract well-observed ICMEs with sheath regions, reflect on whether the spacecraft crosses near the CME nose or flank, and on the background solar wind conditions. Consequently, apparent differences in sheath onset sharpness, compression, or turbulence may partly reflect variations in sampling geometry rather than intrinsic physical differences between events. This observational bias is a well-known limitation of single-point measurements \citep[e.g.,][]{Kilpua2019} and should be considered when interpreting our statistical trends and inter-cycle comparisons.

\begin{table}[!htbp]
\centering
\caption{Summary of Percentage Changes in Median, IQR, and Turbulence Index (TI) for Sheath-Related Variables Across Solar Cycles}
\label{tab:summary}
\resizebox{\columnwidth}{!}{
\begin{tabular}{|c|c|c|c|c|}
\hline
\textbf{Variable} & \textbf{Cycle} & \textbf{Median (\%)} & \textbf{IQR (\%)} & \textbf{TI (\%)} \\
\hline
\multicolumn{5}{|c|}{\textbf{Full Cycle Sample (FCS)}} \\
\hline
$p_{T}$     & SC23 to SC24 & \textcolor{red}{-42} & \textcolor{red}{-51} & \textcolor{red}{-16} \\ \hline
$B$         & SC23 to SC24 & \textcolor{red}{-26} & \textcolor{red}{-36} & \textcolor{red}{-14} \\ \hline
$NRF$ & SC23 to SC24 & \textcolor{blue}{+14} & \textcolor{red}{-36} & \textcolor{red}{-33} \\ \hline
$NRB_{TN}$ & SC23 to SC24 & \textcolor{blue}{+6}  & \textcolor{blue}{+13} & \textcolor{blue}{+6} \\
\hline
$NRB_{TR}$ & SC23 to SC24 & \textcolor{blue}{+6}  & \textcolor{blue}{+22} & \textcolor{blue}{+15} \\
\hline
$NRB_{NR}$ & SC23 to SC24 & \textcolor{blue}{+9}  & \textcolor{blue}{+28} & \textcolor{blue}{+17} \\
\hline
\multicolumn{5}{|c|}{\textbf{Rising Phase Sample (RPS)}} \\
\hline
$p_{T}$     & SC23 to SC24 & \textcolor{red}{-43} & \textcolor{red}{-57} & \textcolor{red}{-25} \\
            & SC23 to SC25 & \textcolor{red}{-37} & \textcolor{red}{-55}  & \textcolor{red}{-28} \\
            & SC24 to SC25 & \textcolor{blue}{+12} & \textcolor{blue}{+6} & \textcolor{red}{-5} \\ \hline
$B$         & SC23 to SC24 & \textcolor{red}{-28} & \textcolor{red}{-42} & \textcolor{red}{-19} \\
           & SC23 to SC25 & \textcolor{red}{-21} & \textcolor{red}{-48}  & \textcolor{red}{-34} \\
            & SC24 to SC25 & \textcolor{blue}{+10} & \textcolor{red}{-10} & \textcolor{red}{-18} \\ \hline
$NRF$ & SC23 to SC24 & \textcolor{blue}{+24} & \textcolor{blue}{+2} & \textcolor{red}{-18} \\
           & SC23 to SC25 & \textcolor{red}{-47} & \textcolor{red}{-49}  & \textcolor{red}{-5} \\
            & SC24 to SC25 & \textcolor{red}{-57} & \textcolor{red}{-50} & \textcolor{blue}{+17} \\ \hline
$NRB_{TN}$ & SC23 to SC24 & \textcolor{blue}{+11} & \textcolor{blue}{+9} & \textcolor{red}{-2} \\
            & SC23 to SC25 & \textcolor{blue}{+1} & \textcolor{blue}{+48}  & \textcolor{blue}{+47} \\
            & SC24 to SC25 & \textcolor{red}{-9}  & \textcolor{blue}{+36} & \textcolor{blue}{+49} \\ \hline
$NRB_{TR}$ & SC23 to SC24 & \textcolor{blue}{+0} & \textcolor{blue}{+34} & \textcolor{blue}{+34} \\
            & SC23 to SC25 & \textcolor{blue}{+4} & \textcolor{blue}{+43}  & \textcolor{blue}{+3} \\
            & SC24 to SC25 & \textcolor{blue}{+4}  & \textcolor{blue}{+7} & \textcolor{blue}{+38} \\ \hline
$NRB_{NR}$ & SC23 to SC24 & \textcolor{blue}{+8} & \textcolor{blue}{+27} & \textcolor{blue}{+18} \\
            & SC23 to SC25 & \textcolor{blue}{+31} & \textcolor{red}{-4}  & \textcolor{red}{-37} \\
            & SC24 to SC25 & \textcolor{blue}{+21}  & \textcolor{red}{-24} & \textcolor{red}{-26} \\ \hline
\end{tabular}}
\end{table}

\begin{table}[!htbp]
\centering
\caption{Summary of Percentage Changes in Median, IQR for global solar wind-Related Variables Across Solar Cycles}
\label{tab:summary2}
\resizebox{\columnwidth}{!}{
\begin{tabular}{|c|c|c|c|}
\hline
\textbf{Variable} & \textbf{Cycle} & \textbf{Median (\%)} & \textbf{IQR (\%)} \\
\hline
\multicolumn{4}{|c|}{\textbf{Full Cycle Sample (FCS)}} \\
\hline
OSF         & SC23 to SC24 & \textcolor{red}{-31} & \textcolor{red}{-14} \\ \hline
\multicolumn{4}{|c|}{\textbf{Rising Phase Sample (RPS)}} \\
\hline
OSF & SC23 to SC24 & \textcolor{red}{-34} & \textcolor{blue}{+0}  \\
           & SC23 to SC25 & \textcolor{red}{-42} & \textcolor{red}{-16}  \\
            & SC24 to SC25 & \textcolor{red}{-12} & \textcolor{red}{-16} \\ \hline
\end{tabular}}
\end{table}

\section{Conclusions} \label{sec:conclusions} 
In this study, we place the ICME sheath region of more than 900 events in the context of the ambient solar wind, examining how sheath characteristics reflects local and global heliospheric solar wind conditions, and how they change over various solar cycles (SC23--SC25). Our main findings for strong ICMEs driving distinct sheath regions are:

\begin{itemize}
\item There is no significant correlation found between OSF and sheath properties, suggesting that the sheath is not directly governed by large-scale heliospheric magnetic flux.
\item A moderate correlation is derived between the sheath's average total magnetic field and the upstream magnetic field (1-hour and 2-hours  ahead), indicating that local upstream conditions do exert a measurable influence on the magnetic field strength within the sheath.
\item Compared to SC23 and 25, the weak SC24 leads to a significant decrease in the sheath's $p_T$ and $B$, as well as OSF, accompanied by a generally reduced variability and turbulence in that variables.
\item $B_N$ increases in dominance moving from upstream into the sheath; most striking for SC25, presumably covering more low-inclination ICMEs, where compression can be achieved more effectively.
\item A transition from tangentially dominated plasma flows in SC23 and SC24 to more normal-oriented flows in SC25 may reflect changes in ICME orientation, which in turn seem to influence the degree of magnetic turbulence and complexity within the sheath.
\item Up to 1 AU both propagation and
expansion mechanisms contribute to sheath formation with a dominance in expansion found for the rising phase of SC25.
\end{itemize}

This work reinforces the importance of ICME sheath regions as dynamic and evolving structures that respond greatly to local solar wind conditions. The observed trends underscore the need to treat sheath evolution as a distinct process, with implications for space weather forecasting and heliospheric modeling. We note though that apparent differences in sheath onset, compression, or turbulence may partly reflect variations in sampling geometry rather than intrinsic physical differences between events. While our statistical approach mitigates this by aggregating a large number of events, future multi-spacecraft missions such as Solar Orbiter and Parker Solar Probe will be essential to resolve the full three-dimensional structure and evolution of sheath regions.

\section{Acknowledgements}
We would like to thank the anonymous referee for constructive comments that helped to improve that work. C.L. acknowledge the support by project PID2024-162117OB-I00 funded by MICIU/AEI/10.13039/501100011033/FEDER, UE. M.O. was part-funded by Science and Technology Facilities Council (STFC) grant number UKRI1207 and Natural Environment Research Council (NERC) grant number NE/Y001052/1.

\bibliographystyle{apalike}
\bibliography{references}

\end{document}